# Innovative static spectropolarimeter concept for wide spectral ranges: tolerancing study

Martin Pertenais*[a,b], Coralie Neiner[a], Laurent Parès[b,c] and Pascal Petit[b,c]

[a]LESIA, Observatoire de Paris, PSL Research university, CNRS, Sorbonne Universités, UPMC Univ. Paris 06, Univ. de Paris-Diderot, Sorbonne Paris Cité, 5 place Jules Janssen, 92190 Meudon, France
[b]Université de Toulouse ; UPS-OMP ; IRAP Toulouse, France
[c]CNRS ; IRAP ; 14 avenue Edouard Belin F-31400 Toulouse, France
*martin.pertenais@irap.omp.eu

## ABSTRACT

Developing an efficient and robust polarimeter for wide spectral ranges and space applications is a main issue in many projects. As part of the UVMag consortium created to develop UV facilities in space (e.g. the Arago mission proposed to ESA), we are studying an innovative concept of polarimeter that is robust, simple, and efficient on a wide spectral range. The idea, based on the article by Sparks et al. (2012), is to use polarization scramblers to create a spatial modulation of the polarization. Along the height of the wedges of the scramblers, the thickness of the birefringent material crossed by the light, and thus the retardance, vary continuously. This variation creates an intensity modulation of the light related to the entrance polarization state. Analyzing this modulation with a linear polarizer, and dispersing the light spectrally in the orthogonal spatial direction, enables the measurement of the full Stokes vector over the entire spectrum. This determination is performed with a single-shot measurement and without any moving parts in the system.

After a quick introduction to the concept and optical design, this article presents the tolerancing study of the optical bench using this spectropolarimeter. The impact of different error sources, such as, birefringence uncertainty or decenter of the wedges, is investigated.

## 1. INTRODUCTION

The concept described in this proceeding is explained in more detail in the articles by Sparks et al. (2012) [1] and Pertenais et al. (2015) [2].

### 1.1 General Theory

To create a static spectropolarimeter, we chose to modulate the polarization of the light in a spatial direction, orthogonal to the main dispersion direction of the spectrograph. To do so, we use birefringent wedges of magnesium fluoride ($MgF_2$) as retardation plates, combined with a linear polarizer [2]. The thickness of the retardation plates varies linearly along its height and, therefore, the retardance imposed to the light also varies linearly along the wedge. Instead of having a given number of sub-exposures and a fixed number of polarization states as for classical rotating modulators, here we have a continuous variation of the polarization state recorded in a single exposure [3].

Using the system described in [2], the final retardance equation is the following, with $\Delta n$ the birefringence of the material used ($MgF_2$ in our case), $\xi$ the apex angle of the first wedge, and $x$ the height of the wedges (with $x=0$ on the optical axis):

$$\Phi = 4\pi \frac{\Delta n(\lambda)}{\lambda} \tan \xi \, x \quad (1)$$

Calculations were performed with the Mueller matrices of the components, and the final output intensity equation is determined, as a function of the entrance Stokes parameters, the retardance equation, and the analyzer angle $\theta$, as [2]:

$$I_{out}(x,\lambda) = 0.5 \cdot [I + Q \cdot (\cos\Phi \cos 2\theta - \sin\Phi \sin 2\Phi \sin 2\theta) + U \cdot \cos 2\Phi \sin 2\theta + V \cdot (\cos\Phi \sin 2\Phi \sin 2\theta + \sin\Phi \cos 2\theta)] \quad (2)$$

## 1.2 Simulation and inversion

For simulations, we fix a wavelength range of [450;750] nm and a wedge height of 3 mm (x ∈[-1.5;1.5]). The angle of the analyzer θ is chosen to represent the most general case; i.e. we avoid that the 3 patterns are mathematically orthogonal, or that the efficiency of a particular Stokes parameter is maximized. We thus chose θ=64°. Equation 2 is then computed and we obtain the simulation of the output image. Figure 1 shows, for example, the result for a circularly polarized entrance light (I=1, Q=U=0, and V=1). The horizontal direction represents the main dispersion, i.e. the wavelength, while the vertical one corresponds to the polarization modulation.

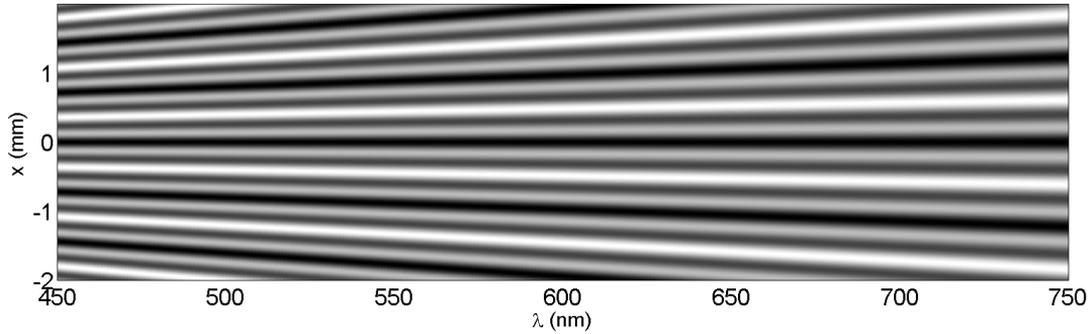

Figure 1 Simulation of the 2D frame obtained with this spectropolarimeter concept. This is an example corresponding to a circularly polarized entrance light, over the visible spectrum.

In practice, we need an inversion code accepting an image of this type as input and computing the value of the Stokes parameters for each wavelength. To do so, Eq. 2 has to be solved with I, Q, U, and V as unknowns and $I_{out}$ as input.
For each wavelength (or every column of the image), this equation is actually overdetermined, as we have n different polarization states (with n the number of pixels in a column), and only four unknowns. Therefore, a classical least-squares solving code is established to solve the system $I_{out}$=A.S at every wavelength, where $I_{out}$ is the input image, A is a vector composed with the trigonometric functions shown in Eq. 2, and S is the Stokes vector we want to determine [2].

## 1.3 Optical design

The core idea of this concept is to illuminate two birefringent wedges with a collimated beam of polychromatic polarized light and to disperse the beam in the orthogonal direction. To create the polarized beam, a polychromatic light source is first collimated and then travels through a rotating polarizer to create linear polarization. An optional quarter waveplate is then used to create –when needed- a circular or elliptical polarization state. The wedges are followed by a second linear polarizer, fixed, used as polarization analyzer. A slit (oriented along the height of the wedges) is then placed in the focal plane of an objective lens to illuminate a grating. Finally, a second objective lens images the spectrum onto a CCD detector, as shown in Fig. 2.

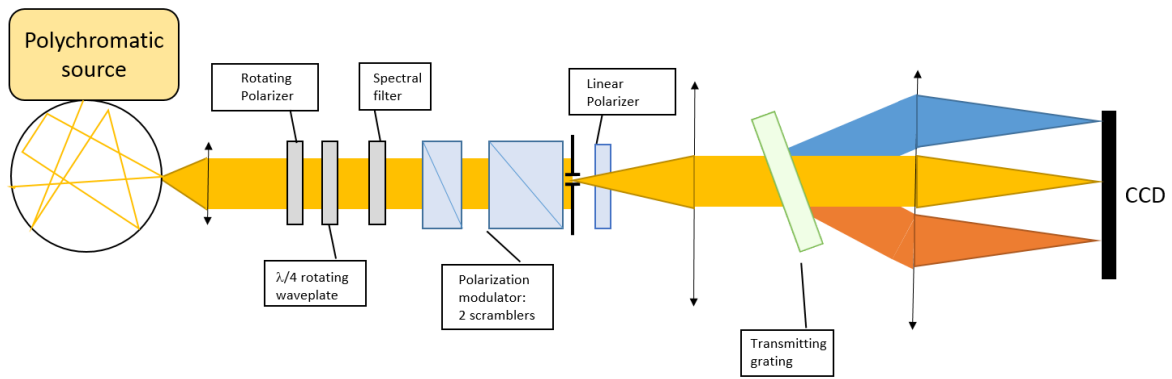

Figure 2 Possible optical design of laboratory tests to prove the validity of the concept of spectropolarimeter presented in this article.

Some laboratory measurements were performed to test the validity of this concept. The detail of the optical bench and the results are presented in the article Pertenais et al. (2015) [2].

## 2. TOLERANCING STUDY

For the inversion process, the theoretical model is used (see Eq. 2). This causes errors in the extracted values, because measured observations show small differences from theory. To calibrate correctly these differences and optimize the quality of the measurements of the Stokes parameters, the various tolerances of the components have to be studied. In this section, the errors added because of birefringence uncertainty, of the precision of the analyzer angle, of the decenter of the wedges, and of the uncertainty on the apex angle of the wedges are detailed.

### 2.1 Error matrix
The accuracy of a polarization measurement can be related to the error matrix of the polarimeter. It is the difference between the real Stokes vector we want to measure (the one of a star for instance) and the actual Stokes vector we determine. The value of the accuracy corresponds to how close our measurement is to the real polarization state. To perform the tolerancing study, the error matrix is used. In theory, the error matrix $\Delta X$ is equal to the null matrix; but adding some uncertainties makes some elements become non-zero. It is then very helpful to identify the various impacts on the measurement.

Non-zero elements on the first line of the error matrix cause depolarization effects. In this case, 100% Q polarization input will not be measured as fully polarized but only 80% Q if the value Q→I is 20% (see Fig. 3). The first column of the error matrix corresponds to induced polarization, this means that polarized light is measured when unpolarized light is actually being observed. The rest of the elements of the matrix are cross-talk (polarization rotation is a particular kind of cross-talk), i.e. cases where wrong polarization states are measured.

This tool is applied below to the various sources of error of this system.

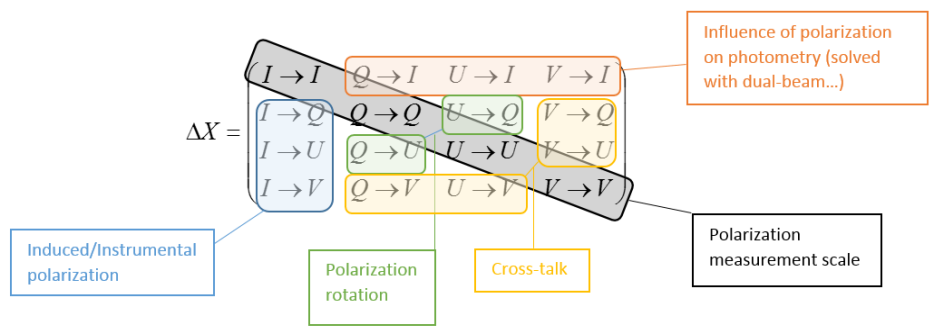

Figure 3 Description of the error matrix to explain the polarization accuracy for tolerancing issues.

To perform the tolerancing study, we make a value vary between its uncertainty range with 50 different configurations. These 50 configurations are computed for 4 different input polarization states: fully unpolarized and fully Q, U and V polarized.

### 2.2 Birefringence uncertainty
A first error using the theoretical model to demodulate the Stokes parameters is due to the value of the birefringence. Indeed, this value is well known theoretically [4][5][6], but changes slightly with the temperature and even with every sample of crystal. For tolerancing issues, we can consider that the change in birefringence is achromatic and that $\Delta n(\lambda) = \Delta n_{th}(\lambda) + \delta \Delta n$. We consider an uncertainty of the value of the birefringence of $\pm$ 5%. As the birefringence of MgF$_2$ is around 0.0118 in the visible range, we use $\delta \Delta n = \pm\, 6.10^{-4}$.

To obtain the error matrix due to birefringence uncertainties, we make the value of $\Delta n$ vary between $\pm\, 6.10^{-4}$ around the nominal value (we compute 50 different configurations within the uncertainty range) and try to demodulate our measurement using the theoretical nominal value. We perform this for four input polarization states (unpolarized, and fully Q, U, and V polarized) and obtain the following error matrix:

$$\Delta X_{\Delta n} = \begin{pmatrix} - & 0.1 & 0.1 & 0.2 \\ 0.05 & - & 0.1 & 0 \\ 0.05 & 0.1 & - & 0 \\ 0 & 0 & 0 & - \end{pmatrix}$$

To conclude, an uncertainty of ±5% on the value of the birefringence imposes various errors on the measurement: a polarization rotation of up to 10%, a depolarization of the linear states of 10%, and a depolarization of circular polarized light of up to 20%.

## 2.3 Analyzer angle precision

The angle of the analyzer placed after the wedges is known with a precision better than ±1°. Even with this large uncertainty, the study shows that its influence can be completely neglected, as the elements of the error matrix are on the order of magnitude of $10^{-5}$.
This uncertainty is thus a negligible error source on the measurement.

## 2.4 Decenter

Another issue can be the relative alignment of the two wedges in the polarimeter. The key point in this concept is the relative difference in thickness at a given height x. Therefore, if the two wedges are decentered with respect to each other, the measurement will be completely wrong. Figure 4 describes the chosen configuration, with d the total height of the wedges and δx the relative decenter.

The total retardance imposed by the first block of wedges remains the same as in theory (as there is no absolute decenter of the first wedge). However, the calculation has to be redone for the second block:

$$\Phi_2(x,\lambda) = \frac{2\pi}{\lambda} x \tan 2\xi\, \Delta n(\lambda) - \frac{2\pi}{\lambda} \tan 2\xi\, \Delta n(\lambda)(d-x) = \frac{4\pi}{\lambda} \tan \xi\, \Delta n(\lambda)(2x-d) \qquad (3)$$

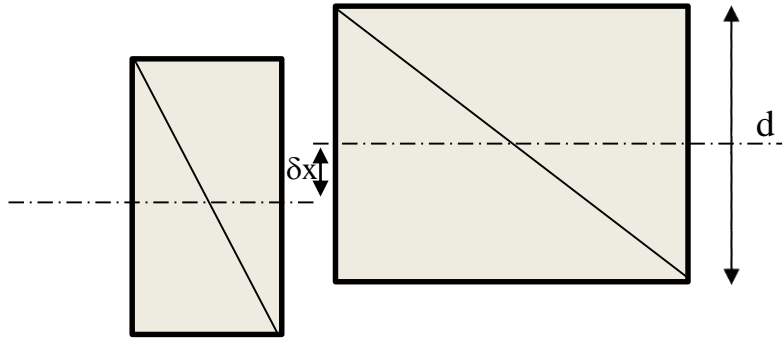

Figure 4 Configuration of the wedges to study the influence of a relative decenter δx on the measurement.

Equation 3 considers that the boundary condition is x=0 at the top the wedge. In theory, we should change the variable to have x=0 on the optical axis. However, as the two optical axes are different in the case we are considering, we change the variable to have x=0 on the optical axis of the first wedge. In practice, $x \leftrightarrow x - \delta x + d/2$

Finally, we get:

$$\Phi_2(x,\lambda) = \frac{4\pi}{\lambda} \tan \xi\, \Delta n(\lambda)\left(2\left(\frac{d}{2} - \delta x + x\right) - d\right)$$
$$\Phi_2(x,\lambda) = \frac{8\pi}{\lambda} \tan \xi\, \Delta n(\lambda)(x - \delta x) \qquad (4)$$

In the Mueller matrix calculations (described in [2]), this retardance is expressed as a function of the retardance of the first block $\Phi_1$. In theory, $\Phi_2=2\Phi_1=2\Phi$. In the present case, we have $\Phi_2=2\Phi-\delta\Phi$, with:

$$\delta\Phi = \frac{8\pi}{\lambda}\tan\xi\,\Delta n(\lambda)\delta x \quad (5)$$

These calculations allow us to express the relative decenter of the wedges as an additional phase retardance to the theoretical one. Using these equations of retardance, we can apply the same process for the tolerancing and determine the influence of a given decenter on the measurement. The following error matrix was calculated for a relative decenter of $\delta x = \pm\,0.05$ mm.

$$\Delta X = \begin{pmatrix} - & 0.1 & 0.2 & 0.1 \\ 0 & - & 0 & 0.3 \\ 0 & 0 & - & 0.1 \\ 0 & 1 & 0.05 & - \end{pmatrix}$$

The impact of the decenter is very important on the measurement. There is, for example, a 100% cross-talk from Q to V, and 30% the other way around.

### 2.5 Apex angle uncertainty

Both wedges used to build the polarimeter module have a given apex angle. The uncertainty given by the manufacturer is $\pm\,0.25°$. While we still use the theoretical angle to demodulate, we make the apex angle of the wedges vary in the simulations. In this way, as for the others uncertainties above, we determine the influence of the uncertainty on the apex angle of the wedges.

Figure 5 shows the results for 50 different configurations within the uncertainty range for a V polarization state as input. After demodulation, we do not see any polarization cross-talk (V↔Q or V↔U), but the depolarization is critical. Indeed, the measured value of V varies between 0 and 1. This means that the circular depolarization due to apex angle uncertainty can be up to 100%.

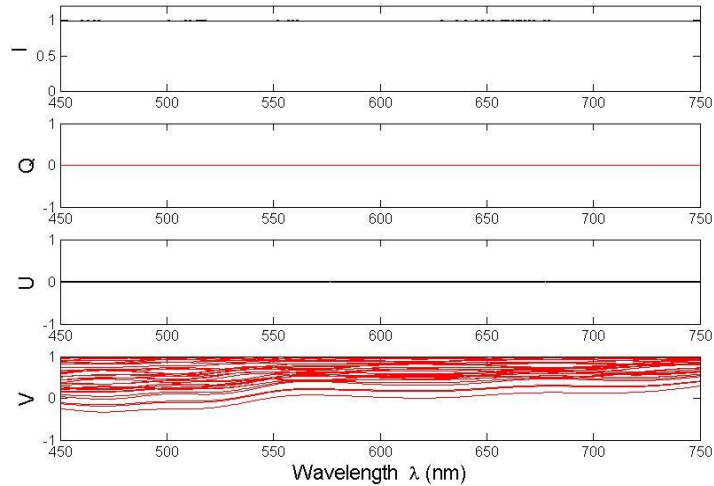

Figure 5 Example of a plot of 50 configurations of apex angle, within the uncertainty range, for a V polarization state input. The depolarization is up to 100%.

As described in the error matrix below, the other errors in the measurement due to apex angle uncertainty are up to 40% depolarization of linear states, up to 20% of polarization rotation, and up to 10% instrumental linear polarization.

$$\Delta X_{\delta\xi} = \begin{pmatrix} - & 0.4 & 0.4 & 1 \\ 0.1 & - & 0.2 & 0 \\ 0.1 & 0.2 & - & 0 \\ 0 & 0 & 0 & - \end{pmatrix}$$

## 3. CONCLUSIONS

This concept of spectropolarimeter offers many advantages. It is indeed completely static, which makes it robust, and compact, as it only needs two small birefringent wedges followed by a linear polarizer. The Stokes parameters are easily extracted from the 2D frames obtained using this concept, by computing a least-squares inversion for every wavelength.

The tolerancing study we performed showed that some errors can occur in the measurement. All of the considered errors combined together give a total error matrix. Of course, this is calculated with various configurations within the uncertainties and thus corresponds to the worst case we could have in reality. However, this total error matrix allows us to extract some interesting features. In particular, we find that the cross-talk, between linear and circular polarization is exclusively caused by decenter of the wedges (more generally, phase retardance error). This fact can be used to help calibrating the system. By adjusting the value of the decenter in the inversion code to suppress the cross-talk between Q and V for example, we can find the actual value and use it instead of the theoretical one (0 in this case), i.e. this error can be calibrated to suppress it. Another fact is that a polarization rotation is mainly due to the apex angle uncertainty, such as the depolarization of circular polarization. The same process as for cross-talk could lead to find the real value of the angles and use them instead of theoretical ones. Another solution to solve the depolarization problem (basically make the first line of the error matrix zero), is to use a dual-beam system. We can replace the linear polarization analyzer by a Wollaston prism, for example, and observe both orthogonal polarization states. Doing this also enables us to solve the photometric issues.

More generally, our study shows that these errors have to be taken into account in the deconvolution process in order to obtain more accurate results. Even if this concept is very sensitive in tolerancing, especially in the alignment of the elements and in the precision and quality of manufactured components, calibrations can be implemented to correct the measurements. The results of the measurements obtained using this concept are available in the article Pertenais et al. (2015) [2]. Therefore, the proposed spectropolarimeter concept appears as a viable option for space missions.